# Diurnal and Seasonal variations of Gravity Waves in the lower atmosphere of Mars as observed by Insight

**J. Hernández-Bernal[1], A. Spiga[1], A. Chatain[2], J. Pla-García[3], D. Banfield[4]**

[1]Laboratoire de Météorologie Dynamique, Sorbonne Université, Paris, France

[2]Laboratoire Atmosphères Observations Spatiales/Institut Pierre-Simon Laplace (LATMOS/IPSL), Université Paris-Saclay, Université de Versailles Saint-Quentin-en-Yvelines (UVSQ), Sorbonne Université, Centre National de la Recherche Scientifique (CNRS), Guyancourt, France

[3]Centro de Astrobiología (CAB), CSIC-INTA, Madrid, Spain

[4]NASA Ames Research Center

Corresponding author: Jorge Hernández Bernal

jorge.hernandez-bernal@lmd.ipsl.fr

**Key Points:**

- Observed gravity waves peak after sunrise and sunset. They are almost absent during aphelion season, and more prominent around equinoxes.
- Observed waves often exhibit phase coherence across different sols and coincidence in time with increments of absolute pressure.
- The overlap between gravity waves and high-order harmonics of thermal tides is explored.

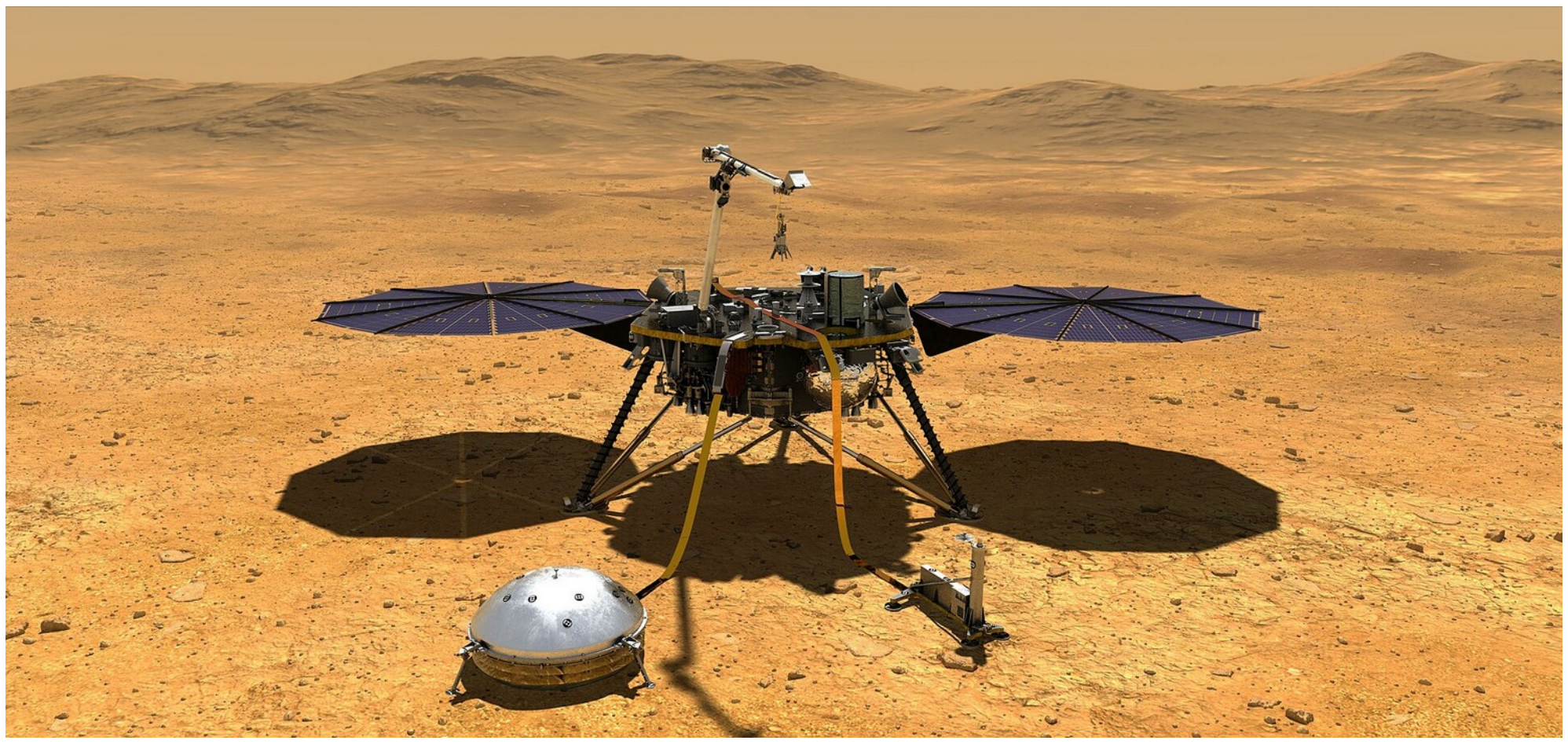

## Abstract

We investigate Gravity Waves (GWs) in the lower atmosphere of Mars based on pressure timeseries acquired by the InSight lander. We compile a climatology showing that most GW activity detected at the InSight landing site takes place after the sunrise and sunset, they are almost absent during the aphelion season, and more prominent around the equinoxes, with variations during dust events and interannual variations. We find GWs with coherent phases in different sols, and a previously unnoticed coincidence of GW activity with those moments in which the diurnal cycle (of tidal origin) exhibits the fastest increases in absolute pressure. We explore the possibility that some of these GWs might actually be high-order harmonics of thermal tides transiently interfering constructively to produce relevant meteorological patterns, and discuss other interpretations based on wind patterns. The so-called Terminator Waves observed on Earth might also explain some of our observations.

## Plain Language Summary

Gravity waves are oscillations in a fluid, observed for example in the surface of water when it is perturbed. These waves are common in planetary atmospheres and they play a key role in their dynamics, but they are difficult to reproduce in computer models. Based on the comprehensive observations of pressure acquired by the InSight lander, we analyze the occurrence of these waves on Mars at different seasons and times of the day with unprecedented detail. We find that these waves are much more common after the sunrise and the sunset. Also, they tend to repeat synchronously in consecutive days, and they seem to happen at those moments in which the absolute atmospheric pressure is increasing. Interpreting these observations is challenging, and we propose various plausible explanations.

## 1 Introduction

Gravity Waves (GWs) in a fluid are waves whose restoring force is buoyancy. In a stably stratified atmosphere, they can propagate through long distances both in horizontal and vertical directions, transporting energy and momentum that can be deposited by wave breaking, which has an effect on the global dynamics of planetary atmospheres. Nappo (2013), and Fritts and Alexander (2003) are excellent introductions to the topic.

GWs have been observed on Mars in the vertical profiles acquired by missions descending to the surface (Magalhães et al., 1999; Banfield, Spiga, et al., 2020), orbital missions sensing directly the upper atmosphere (by aerobraking or mass spectrometers, England et al., 2017; Terada et al., 2017; Vals et al., 2019), or using remote sensing instruments, like radioscience experiments (Creasey et al., 2006; Tellman et al., 2013), or spectrometers (Heavens et al., 2020; Nakagawa et al., 2020). Imaged cloud patterns (Pirraglia, 1975; Río-Gaztelurrutia et al., 2023) and pressure fields inferred from hyperspectral imaging (Spiga et al., 2007) from orbit also reveal the presence of GWs. Finally, ground stations have acquired timeseries in which GWs are present in various variables (Haberle et al., 2014; Banfield, Spiga, et al., 2020; Guzewich et al., 2021; Sánchez-Lavega et al., 2022).

Observations from ground stations and satellites each have their advantages and disadvantages. Ground stations can only observe the lowest layers of the atmosphere at a single location, but they can potentially observe all local times and seasons with high accuracy. Satellites present an improved spatial coverage, but their temporal coverage and sensitivity are worse. In addition to it, the sensitivity of different observational techniques differs depending on the spatial and temporal scale of GWs (wavelength and period respectively).

Given the limitations in the spatial and temporal coverage of observations, simulations have the potential to enlighten our global understanding of GWs in the Martian atmosphere. However, the limited grid resolution of usual GCMs (General Circulation Models) limits their capability to resolve GWs, which are often parametrized (e.g. Medvedev et al. 2011; Gili et al., 2020) or explored in mesoscale models (Spiga et al., 2012; Altieri et al., 2012; Guzewich et al., 2021; Sta. Maria et al., 2006) and high-resolution GCMs (e.g. Kuroda et al., 2016; 2019; Kling et al. 2022).

The presence and properties of GWs at a given location and time depends on the local source mechanisms and on the propagation of GWs from other locations. Typical sources of GWs (Fritz and Alexander, 2003) include winds impinging topographic obstacles (i.e. orographic waves; e.g. Hernández-Bernal et al., 2022;

Maria et al., 2006), convection (e.g. Spiga et al. 2013), and wind shear. Orographic GWs can only be generated near the planetary surface, while convection and wind shear GWs can happen at any altitude. The vertical propagation of GWs is conditioned by the filtering produced by different atmospheric layers, which is sensitive to the wavelength of GWs, and the horizontal propagation is also subject to filtering. During the hours of maximum insolation, the near surface atmosphere is characterized by the presence of the Planetary Boundary Layer (PBL; Read et al., 2017), which involves an unstable atmosphere not favorable for GW propagation, and turbulence is the dominant source of pressure variations recorded close to the surface by a lander like InSight (e.g. Chatain et al., 2021). The PBL collapses during the Martian night, and then the atmosphere becomes stable (Read et al., 2017; Banfield et al., 2020).

Climatologies of GWs observed from the ground are not very common, and the most comprehensive one so far is the one compiled by Guzewich et al. (2021), based on the pressure sensor in MSL (Mars Science Laboratory). Climatologies for the middle atmosphere made from satellite data are more common (e.g. Creasey et al., 2006; Heavens et al., 2020). The relation between GWs in the lower and middle atmosphere of Mars remains unexplored (an example of such exploration on Earth can be found in Groot-Hedlin et al., 2017), which is in part due to the fact that many of the observations of GWs in the middle atmosphere are made by sun-synchronous orbiters that only observe a few local times, while landers can observe at all local times but are constrained to a specific location.

In addition to the aforementioned mechanisms that usually produce GWs, the movement of the terminator could also produce GWs. The so-called Terminator Waves (TWs) consist, according to the literature, of GWs generated around the terminator by the sudden change of insolation.TWs have been reported on Earth and might also exist on Mars. Beer (1973) first suggested that the fast displacement of the terminator and the consequent change of insolation, could produce GWs around the terminator on Earth, and on Mars (Beer, 1974). Different authors have reported observations on Earth identified as TWs in the ionosphere (Raitt and Clark, 1973), the upper neutral atmosphere (Forbes et al., 2008), and the troposphere (Hedlin et al., 2018). Miyoshi et al. (2009) investigated the observations of TWs by Forbes et al. using a GCM. To our knowledge, only Forbes & Moudden (2009) have investigated TWs on Mars; they reported the reproduction in a GCM of a TW happening in latitudes 30ºS-10ºN after the sunset at an altitude of 160 km, and extending to lower altitudes.

In this paper, we benefit from the unprecedented quality in terms of temporal coverage and accuracy, of the pressure timeseries acquired by the Pressure

Sensor (PS) on the InSight Lander (Banfield, Spiga, et al., 2020; Banerdt et al., 2020; Banfield et al., 2019; Spiga et al., 2018). Insight (Interior Exploration using Seismic Investigations, Geodesy, and Heat Transport)  landed on Mars on November 26, 2018, and it operated for 1440 Martian sols, up to the end of 2022. The PS acquired continuous measurements during most of the first ~800 sols with a noise level of 50mPa at frequencies below 0.01 Hz (Banfield, Spiga, et al., 2020).

The mostly continuous temporal coverage is particularly outstanding when compared to other landers, which typically observe only at specific times each Martian sol. This enables us to compile a new climatology of GWs and explore some of their features. Since we only use pressure data, we cannot ensure that all the oscillations observed are actually GWs, but for simplicity we refer to them as GWs in all cases. Banfield, Spiga, et al. (2020) combined the PS with the TWINS (Temperature and Wind for InSight) wind sensor to confirm that some of these pressure oscillations were indeed GWs.

In this paper we explore some potential relations between GWs and thermal tides. Thermal tides are planetary-scale oscillations (that can be seen as a particular kind of a gravity wave) consistent on globally coherent oscillations with periods that are precisely harmonics of the solar day (harmonic 1, 24h; harmonic 2, 12h, harmonic 3, 8h, and so on). Non-tidal GWs are expected to be transient local or regional oscillations  with periods usually larger than a couple of minutes (the Brunt Väisälä frequency acting as an upper limit for GW frequency) and shorter than one hour, or shorter than a few hours in some cases.

In a recent paper (Hernández-Bernal et al., 2024; HB24 from now onwards), we showed that high-order tidal harmonics even beyond harmonic 24 could be present in the atmosphere of Mars based on the PS. This finding hints at a region in the spectrum of atmospheric oscillations (to which we will refer as “the gray zone”) in which both thermal tides and GWs could coexist. We suggested that these high-order tidal harmonics can transiently interfere constructively to produce daily repeating meteorological patterns that could even be confused with GWs. Following HB24, in this work we explore the overlap between thermal tides and GWs in the spectrum of oscillations in the Martian atmosphere.

The outline of this paper is as follows. In section 2 we introduce the dataset under study and the methods used in this work. In section 3 we present our climatology of GWs and analyze some features of those GWs. In section 4 we discuss the observations, explore the overlap between GWs and thermal tides in the light of our observations, and consider the possibility that TWs are present in our observations. In section 5 we summarize and give some perspectives.

## 2 Dataset and methodology

All our analyses and data processing scripts are made using python codes, and many of our algorithms rely on the Numpy (Harris et al., 2020) and Scipy (Virtanen et al., 2020) software packages.

### 2.1 The InSight PS dataset

We access the InSight PS dataset through the derived dataset produced by HB24 and available in a public repository (Hernández-Bernal et al., 2023). This is simply a derived dataset with data organized in long intervals of continuous measurements without gaps longer than 70s within them. Many of these intervals cover periods of dozens of sols. This dataset contains 630 full sols with continuous measurements between sol 15 and sol 824 of the InSight mission, both included. It starts at Ls 304º (Solar Longitude 304º) in MY34 (Martian Year 34), and ends at Ls 20º in MY36. Starting at sol 825, sols with continuous coverage adequate for our analysis are sparse and so we do not consider that period.

### 2.2 Nomenclature for seasons and Martian timekeeping

In the context of this paper we use the nomenclature for meteorological seasons centered on equinoxes and solstices introduced in HB24. The northward equinox season (sun moving northward in the Martian sky) refers to Ls 315º-45º, northern solstice season (sun around its solstitial northernmost position in the Martian sky) refers to Ls 45º-135º, southward equinox season refers to Ls 135º-225º, and southern solstice season refers to Ls 225º-315. We will show that this non-conventional division of seasons is relevant for the description of our results.

The Local Mean Solar Time (LMST) is usually used in the operations and archiving of data from meteorological stations on the ground of Mars. We will use in this paper the Local True Solar Time (LTST), which is more convenient for the analysis of data and comparison to the models because it is directly connected to insolation. From now onwards we will simply use LT (Local Time) to refer to LTST.

Because InSight is very close to the equator (latitude 4.5ºN), the sunrise and sunset occur respectively at 6 and 18 LT, with variations of less than +/-10 minutes over the year.

Timekeeping calculations are made using algorithms based on Allison et al. (1997, 2000).

## 2.3 Signal processing and analysis

### 2.3.1 Spectral filters

Bandpass, lowpass, and highpass filters enable us to remove very high and very low frequencies from a signal. In the analysis of GWs, it is useful to remove very high frequencies (i.e. infrasounds), and very low frequencies (e.g. thermal tides), or in general, to isolate specific signals. We use subroutines provided by Scipy (Virtanen et al., 2020).

### 2.3.2 Wavelet analysis

Wavelet analysis was previously applied to analyze GWs in data acquired by the PS on Insight by Banfield, Spiga et al. (2020), and in the equivalent dataset obtained from the Rover Environmental Monitoring Station (REMS) onboard the Mars Science Laboratory (MSL) rover Curiosity by Guzewich et al. (2021). For a given one-dimensional time signal, it results in a two-dimensional time-frequency map showing the wavelet power, which indicates the presence of different frequencies at different times in the signal (bottom panel in Figure 1). We use the codes provided by Predybaylo (2014) based on Torrence and Compo (1998). The specific configuration of our wavelet analysis is discussed in the supporting text S1.

Wavelet analysis consists on the correlation of a given wavelet function with the signal; this correlation is evaluated for each instant of time and for different frequencies of the wavelet function. This implies that features of the signal that are not necessarily waves, can produce signals in the wavelet results that don't correspond to actual wave-like signals, this is the case of turbulence, which was already analyzed by Chatain et al. (2021). As a notable example, we find that pressure drops caused by dust devils (analyzed in this dataset by Spiga et al. 2021), given their large pressure anomaly, produce large correlation values at many different frequencies, leading to spurious signals in the wavelet power graphs.

The starting and the end of the time dimension of the wavelet power are affected by border effects and a wider margin for larger periods needs to be considered. In order to avoid this effect, we employ, for each sol of the dataset, a timeseries with a margin of 4 additional hours at the starting and the end. This is not possible for all the sols, which reduces the dataset (see section 2.1 and supporting text S2) from 630 to 615 full sols when it comes to the wavelet analysis.

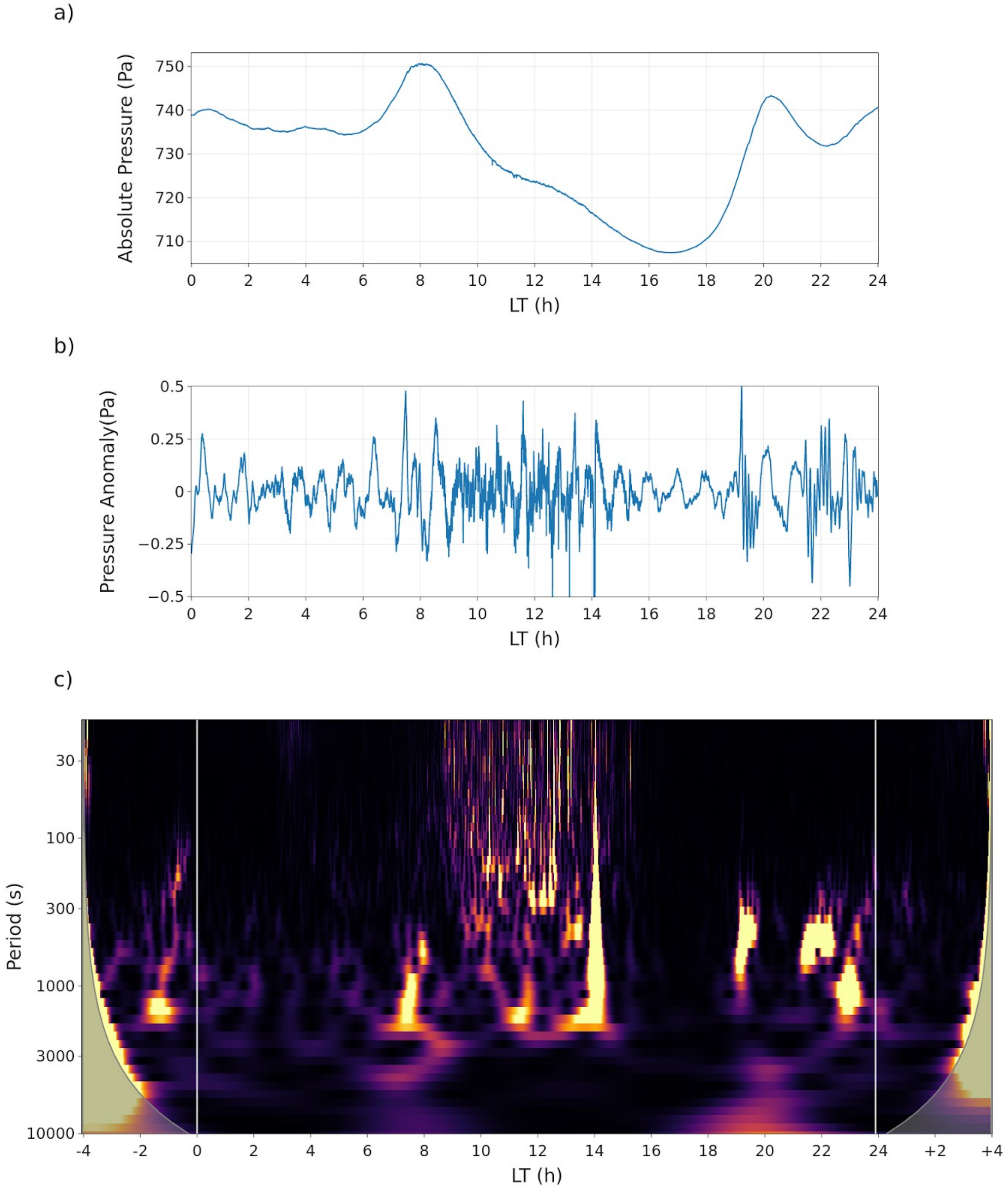

**Figure 1**. Illustration of analysis techniques applied to the InSight dataset to study GWs on sol 129. (top) Original signal, diurnal variations of absolute pressure corresponding mostly to thermal tides are present. (middle) Signal with a highpass filter at a period of 3700s, low-order harmonics of thermal tides and other low frequency oscillations are removed. (bottom) Wavelet power graph, with 4 hours of margin at the end of the previous sol and at the starting of the next sol. During the hours of maximum insolation (9-16 LT) the signal is dominated by turbulence. Clear cases of wave activity are present around 7 LT and from 19 to 24 LT.

Average wavelet power is systematically larger for lower frequencies, which will result in larger wavelet power and an overrepresentation of low frequencies over high frequencies in graphs where wavelet power for different frequencies are represented at the same scale. As a solution, we normalize the wavelet power at each frequency by dividing by its average for the whole dataset. This approach is not free of caveats, for example, it can result in bias if the average signal at a particular frequency is especially strong or weak, and the information about the relative strength of signals at different frequencies is lost, but it provides more convenient representations of the distribution of wave activity.

### 2.4 Context of global dust content

Dust is known to be a key to the dynamics of the Martian atmosphere. As we did in HB24, we discuss our findings on this dataset in the context of the global dust content as computed from observations by Montabone et al. (2015; 2020), and we follow Kass et al. (2016) for naming  annually repeating dust events (A, B, C). Along this paper we will refer to the following dust events (same as in HB24): dust event C in MY34 (sols 40-80; C34 from now onwards), regional off-season dust event in MY35 (sols 180-220; R, for “regional”), dust event A in MY35 (sols 550-650; A35), and dust event C in MY35 (sols 700-740; C35).

The C dust event was much stronger in MY34 compared to MY35. We can expect some interannual variations in this part of our dataset observed in both years due to the different dust concentrations.

## 3 Observations of GWs

This section describes and analyzes the observations of pressure oscillations with periods between 100s and 3700s (one Martian hour). The lower boundary is given by the frontier between GWs and infrasounds, previously estimated to be at 100s by Banfield, Spiga et al. (2020). The higher boundary of one Martian hour is an election motivated by the report of high-order tidal harmonics with periods around one hour in HB24.

We assume that these pressure oscillations correspond to GWs, keeping the caveat in mind that cannot be fully confirmed based only on the observation of pressure. Banfield, Spiga et al. (2020) analyzed pressure oscillations representative of those reported here together with simultaneous wind oscillations, showing that those oscillations corresponded to actual GWs. Unfortunately the wind dataset of InSight is not as complete as the pressure dataset in terms of temporal coverage, and so we choose to rely only on the pressure dataset (a path previously followed by Guzewich et al., 2021).

We start this section by presenting the climatology of GWs as revealed by the wavelet analysis, in section 3.1. In section 3.2 we discuss interannual variations and the effects of dust. In section 3.3 we present the sol-to-sol coherence of waves. Then in section 3.4 we show the coincidence between the observed GWs and the variations of absolute pressure.

### 3.1 The wavelet climatology of GWs

Fig. 2 displays the climatology of wavelet power in terms of sol/Ls, LT, and period. The individual pressure signals corresponding to each part of this climatology can be explored in supporting animations S1, S2, S3. We visually inspected these animations to validate the systematic wavelet analysis in terms of occurrence of waves and detected frequencies. Local times from 9 to 16 LT are shadowed in fig. 2a because this part of the diurnal cycle is dominated by turbulence. Also, Chatain et al. (2021) found nocturnal turbulence during the southern solstice season, and whether or not we find waves or only turbulence at the night time in this season is unclear to us, as exposed in subsection 3.1.3.

Gravity wave activity recorded by InSight exhibits diurnal and seasonal variability. Most wave activity happens in seasons other than the period from Ls 20º to Ls 155º (around the northern solstice at Ls 90º). Other seasonal trends are different at different LTs, and therefore our discussion is organized in terms of LT. Most wave activity takes place in the early morning (6 to 9 LT) and the evening (18 to 24 LT), following the sunrise and the sunset. GWs in the evening following the sunset observed by InSight were already preliminarly reported by Banfield, Spiga et al. (2020). Nocturnal activity (0 to 6 LT) is also quite present. During the hours of maximum insolation (9 to 16LT) the observed wavelet power is basically caused by turbulence, which is out of the scope of this work. And before the sunset (16 to 18 LT) a calm period (previously reported by Banfield, Spiga et al. 2020) takes place regardless of the season.

Wave amplitudes can be observed in supporting animations S1, S2, S3. They are typically around 0.2-0.5 Pa. We analyze a single MY starting at Ls 315º in MY34, and then we analyze interannual variations in section 3.2.

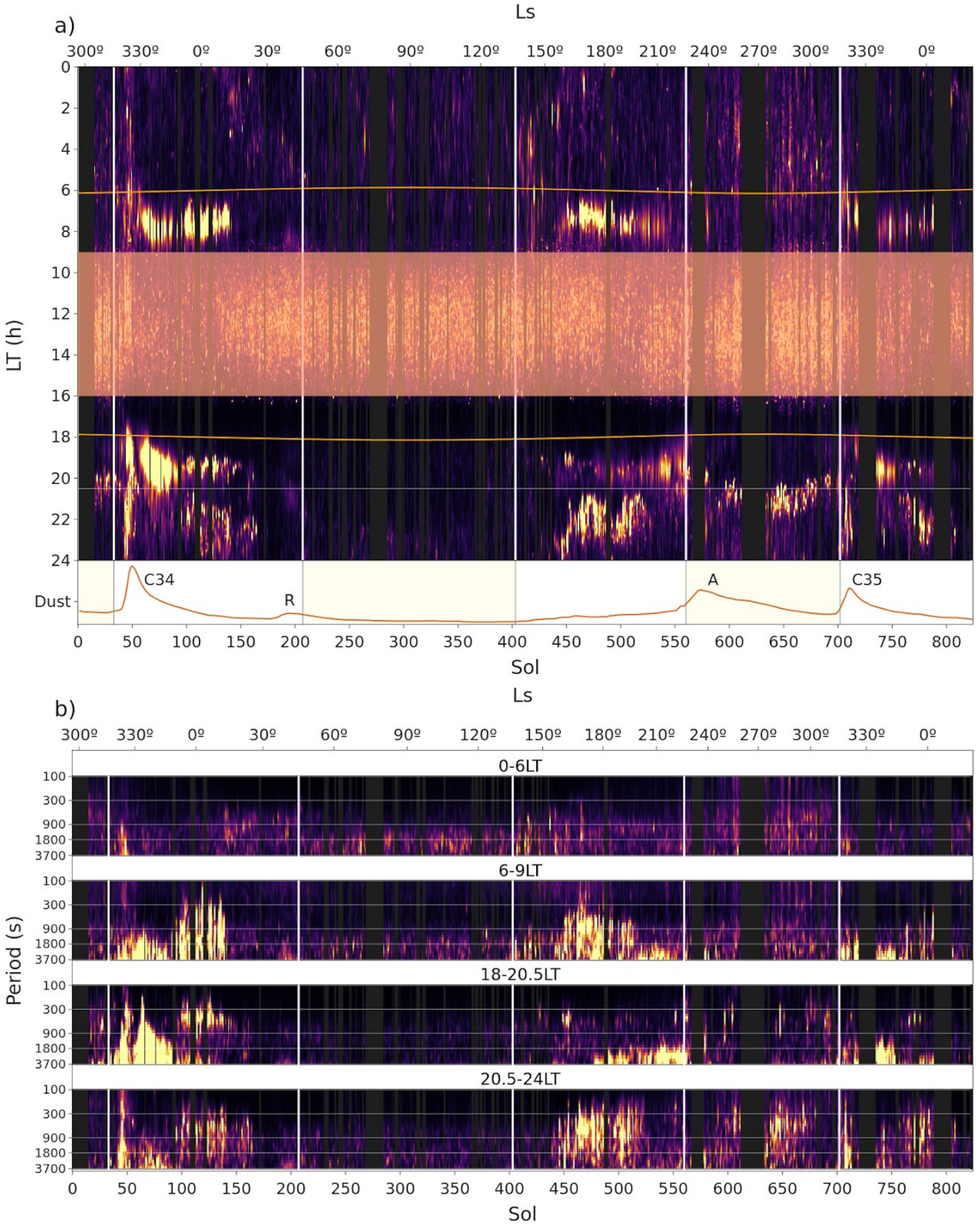

**Figure 2**. Climatology of wavelet power in terms of Ls, LT, and wave period (from 100s to 3700s). Brighter colors correspond to larger wave activity, different periods have been normalized as detailed in section 2.3.2. Dark gray vertical stripes correspond to sols that are not available in our dataset. White vertical lines indicate meteorological seasons as defined in section 2.2. An interactive version of this figure is available in the supporting material. a) Climatology in terms of Ls and LT,

with different periods normalized. Diurnal turbulence from 9 to 16 LT is shadowed by a reddish square. The white horizontal line at 20.5 is for reference (see section 3.1.2), Orange curves around 6 and 18 LT correspond to the sunrise and sunset. The global dust content (section 2.4) is included for quick reference. This figure compares to figures 6 and 9 in Guzewich et al. (2021). Animation S4 is equivalent to this panel separated by wave periods. b) Climatology in terms of Ls and Period for various LT ranges. This graph provides information on the wave periods complementary to panel a. Each graph covers different LT ranges; within each one of those ranges, different local times have been averaged.

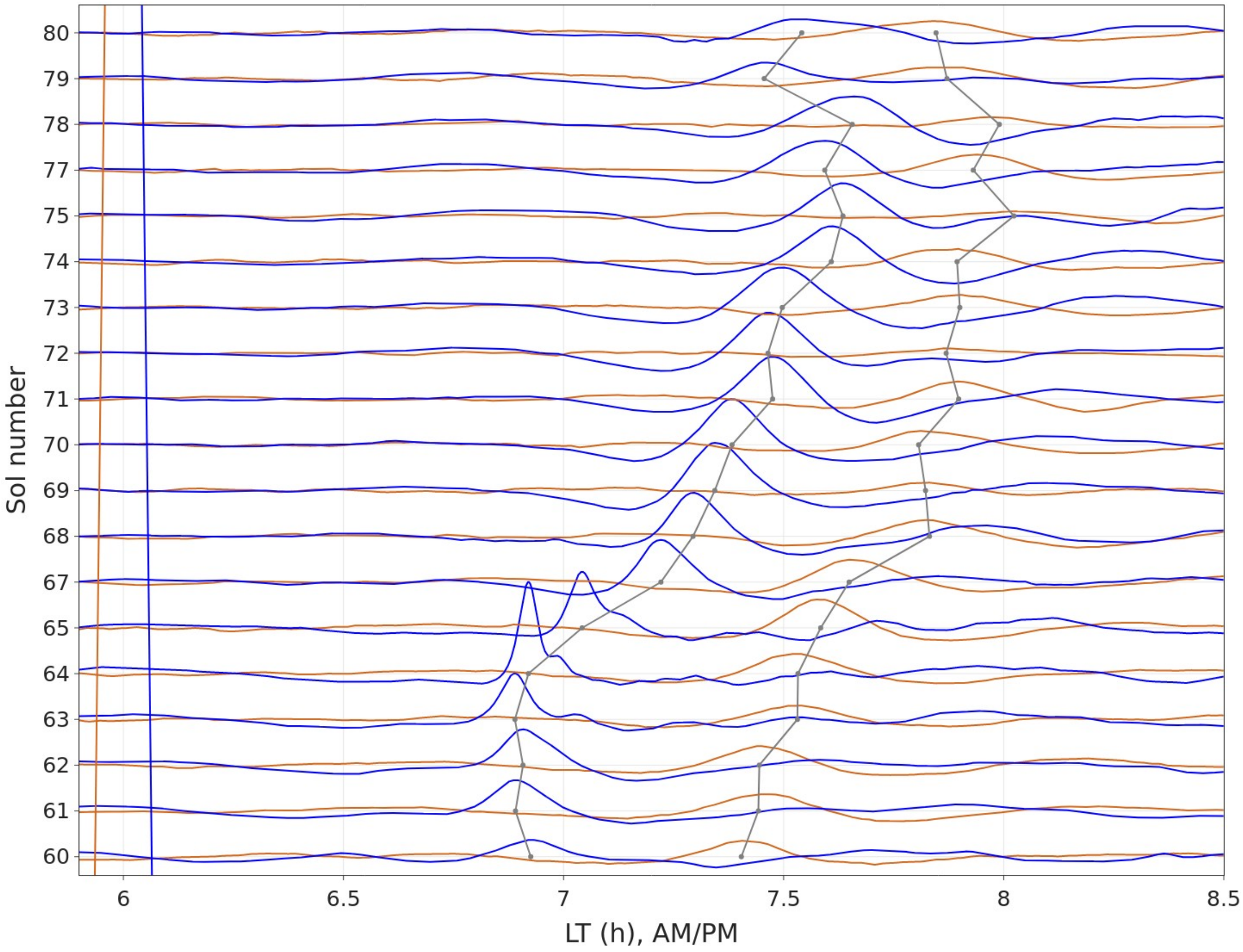

**Figure 3**. Soliton-like signal, analogue to fig. 4b in Banfield, Spiga et al. (2020), but in addition to the signal in the evening (blue), the signal corresponding to the morning has been added (orange). The evening signal reaches higher amplitudes. They are separated by almost exactly 12h, the morning signal is delayed by 0.5h from that. Gray curves indicate the local time of the pressure anomaly maxima (center of the soliton-like signal) for each sol; the shift in local time over sols occurs in parallel. A bandpass filter with window 100s-3700s was applied to the signals, and that reduced their amplitudes. The vertical orange and blue curves in the left indicate the LT of sunrise and sunset respectively. In the horizontal axis, AM/PM stands for "Ante Meridiem"/"Post Meridiem", common terms to refer to the local time before and after noon.

### 3.1.1 Early morning GWs (6-9LT)

Waves in the early morning are observed between the sunrise and the starting of the turbulence, from 6 to 9 LT. Most wave activity takes place in the northward equinox season up to Ls 15º, and the southward equinox seasons since Ls 155º, at local times centered around 7-8 and evolving smoothly in each season (fig. 2a). Low and mid frequencies (period 900-3700s) prevail, higher frequencies (period 100-900s) are less common and they exhibit a preference for a short season around the equinoxes (Ls 350º-15º, and Ls 160º-200º; fig. 2b).

Interestingly, a soliton-like signal like the one identified by Banfield, Spiga et al. (2020) is present during the same period almost 12h (11.5h) in advance in the morning and showing a similar evolution (fig. 3). Interestingly analogue signals have been occasionally observed on Earth (Tsai et al., 2004).

### 3.1.2 Evening GWs (18-24 LT)

While most wave activity in the morning happens in a single interval around 7-8LT, the distribution in the evening is more complex. As seen in fig. 2a, it appears distributed in one or two horizontal stripes corresponding to different local time intervals, depending on the season, which is the reason why in fig. 2b we have differentiated two LT ranges: 18-20.5LT, and 20.5-24LT. The wave patterns displayed in fig. 4 are a good example of these waves in the evening.

Between the end of the C dust event (Ls ~330º) and Ls 350º, wave activity after 20.5 LT is weaker and constrained to low frequencies, most activity during this short season takes place from 18 to 20.5LT, with wave periods typically longer than 900s. This wave activity from 18 to 20.5 and from Ls 330º to Ls 350º corresponds to the soliton-like signal, also discussed in section 3.3 and displayed in fig. 3. Starting in Ls 350º, two different stripes are present. Both stripes contain waves with higher frequencies (periods between 100s and 900s), and also smaller activity at lower frequencies (periods reaching 3700s). These two stripes were illustrated by Banfield and Spiga (2020), in their fig. 4a, and as they mentioned, no more waves are detected after sol 150 (Ls ~20º).

Following a notably long pause around the northern solstice season, wave activity returns in the southward equinox season, starting at Ls 155º, it appears in fig. 2a the form of two stripes very similar to those of the northward equinox season. The stripe from 18 to 20.5 LT exhibits a small boost of activity at high frequencies (period ~400s) at the starting of the wave season (Ls 155º), and then wave activity at this LT range fades until Ls 175º, when low frequency waves appear, they remain until the end of the southern equinox season (Ls 225º). The stripe from 20.5 to 24 LT displays wave activity with frequencies ranging from 100s to 2000-3000s in Ls 155º-200º. A sample of these waveforms is presented in fig. 4, and it will be further discussed in section 3.3.

During the southern solstice season, a single stripe is apparent in fig. 2a. The LT of this stripe falls mostly in the second LT range (20.5-24), but it exhibits a seasonal evolution and part of it happens in the first range (18-20.5). These waves exhibit periods predominantly in the range 300-900s, but other frequencies are present to a lower extent.

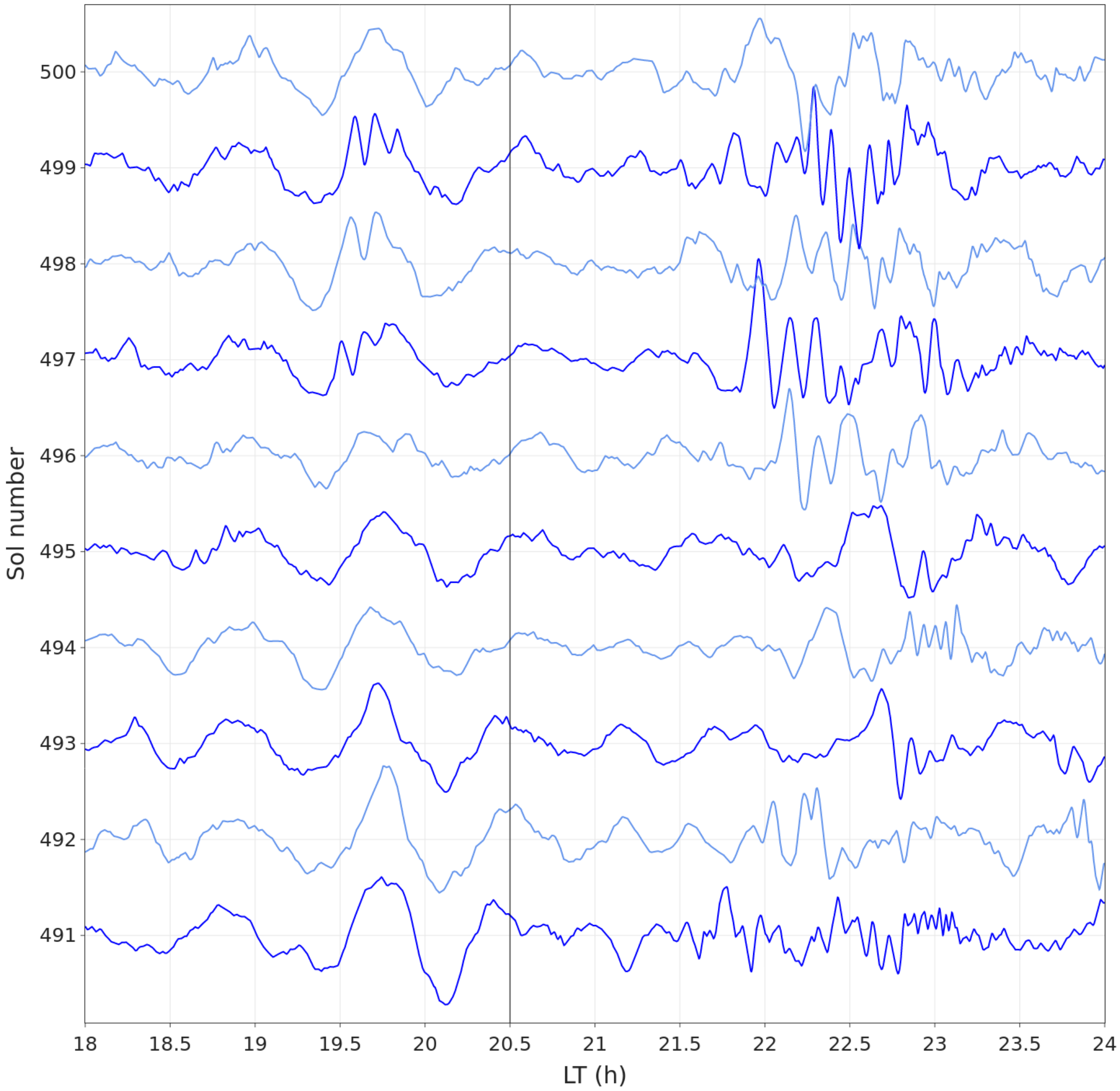

**Figure 4**. Pressure anomalies from sol 491 to sol 500 (Ls 183-187º; southward equinox season), after a bandpass filter from 100s to 3700s. The vertical line at 20.5 LT indicates the different LT ranges. The different colors are simply for clarity. The reader can refer to supporting animations S1-S3 for more detailed graphs for all sols including units.

### 3.1.3 Nocturnal GWs (0-6LT)

Nocturnal wave activity appears weak in fig. 2. However, some good examples of waves can be found at this LT range, for example, around 3LT in sol 140 (available in supporting animation S1).

Nocturnal wave activity is characterized by higher daily variations (lower repeatability), and weaker seasonal trends. On balance, nocturnal waves are less predictable.

Most of the time, nocturnal activity occurs in low and mid frequencies (period>900s). However, activity below 900s boosts during the end of the northward equinox season (Ls 15-45º) and during the starting of the southward equinox season (Ls 135-170º). It is worth noting that this boost in the boundaries of the equinox seasons complements very well with activity in the morning and evening in the same seasons: when waves are more active during the nocturnal period, they are less active during the other LT ranges. The lack of GWs at high frequencies around the northern solstice is in coincidence with the hiatus in nocturnal turbulence reported in this season by Pla-García et al., (2023).

During the southern solstice season, wavelet power is stronger at higher frequencies (period<900s) in fig. 2b. Chatain et al. (2021) found nocturnal turbulence during the same season, which is probably the reason for this observation. While GWs might also be present among this turbulence, we leave that for future investigations.

### 3.1.4 Diurnal Turbulence and calm period (9-18 LT)

From 9 to 16 LT we obtain strong signals, but most of them correspond to turbulence and convective vortex, which were studied by Spiga et al. (2021) and Chatain et al. (2021). While GWs might be present within the diurnal turbulence, those are difficult to disentangle and will not be subject of this paper. Around 16 LT, diurnal turbulence fades and a calm period without much GW activity extending up to 18 LT starts (previously reported by Banfield, Spiga et al., 2020). This corresponds to the hours before sunset, when insolation falls, temperatures start to drop, and the boundary layer collapses. The calm period extends to later local times in the northern solstice seasons.

### 3.2 Effects of dust and interannual variations

Some features in fig. 2a suggest that particular dust events left a footprint that is present in the climatology of wavelet power. In the case of the annually repeating C dust event, it was recorded in two different years, and that enables an interannual comparison. We focus this analysis on dust events C34, C35, and R.

A GDE (Global Dust Event) took place in MY 34 (Sánchez-Lavega et al., 2019) and it altered the global atmosphere compared to non-GDE years. In addition to it, the C dust event later in the same year (C34) was larger than the one in MY35 (C35), as we can see in the dust content curve available in the bottom of fig 2a. Wavelet power was overall larger in the northward equinox season of MY34 compared to that of MY35, which might be the result of the larger C dust event. It is evident both in C34 and C35 that wavelet power was larger at a wider range of local times and it reached higher frequencies (below 300s), compared to other sols in the same season without dust event. Chatain et al. (2021), in their fig. 2, showed signs of nocturnal turbulence during C dust events. The wavelet power observed here probably contains such turbulence, and it might contain GWs.

By looking at fig. 2, we can conclude that the overall trends described in section 3.1 for the northward equinox season were present also in the following year when the dust event C35 faded. This includes the soliton-like signal, which can be found in supporting animations S2 and S3 both in the morning and the evening around sol 648, but it was weaker in MY35 compared to MY34, maybe because of the weaker C dust event.

Apart from C dust events, we note that the R dust event might have also left a footprint in wavelet power. This dust event was a large storm that took place shortly before sol 200, at the end of the northward equinox season, in MY35. Wavelet power was larger in coincidence with this event at 4-5, 8-9, and 20-21 LT, according to fig. 2a. Fig. 2b indicates that this wavelet power corresponds to low frequencies (period 1800-3700).

### 3.3 The sol-to-sol coherence of GWs

GWs are in general expected to exhibit random phases, however, when looking at the bandpassed pressure timeseries we find oscillations that repeat with a similar phase in consecutive sols. Fig. 4 displays some clear examples of these sol-to-sol coherence in sols 491-500 between 18 and 21 LT, with a main peak repeating in all the cases around 19.7LT, and some secondary peaks repeating most sols around 18.8, and 20.6. Some other peaks repeat less regularly. It is easy to find clear examples of these waves in the dataset once a bandpass filter has been applied. Many clear examples can be spotted in supporting animations S1 and S2.

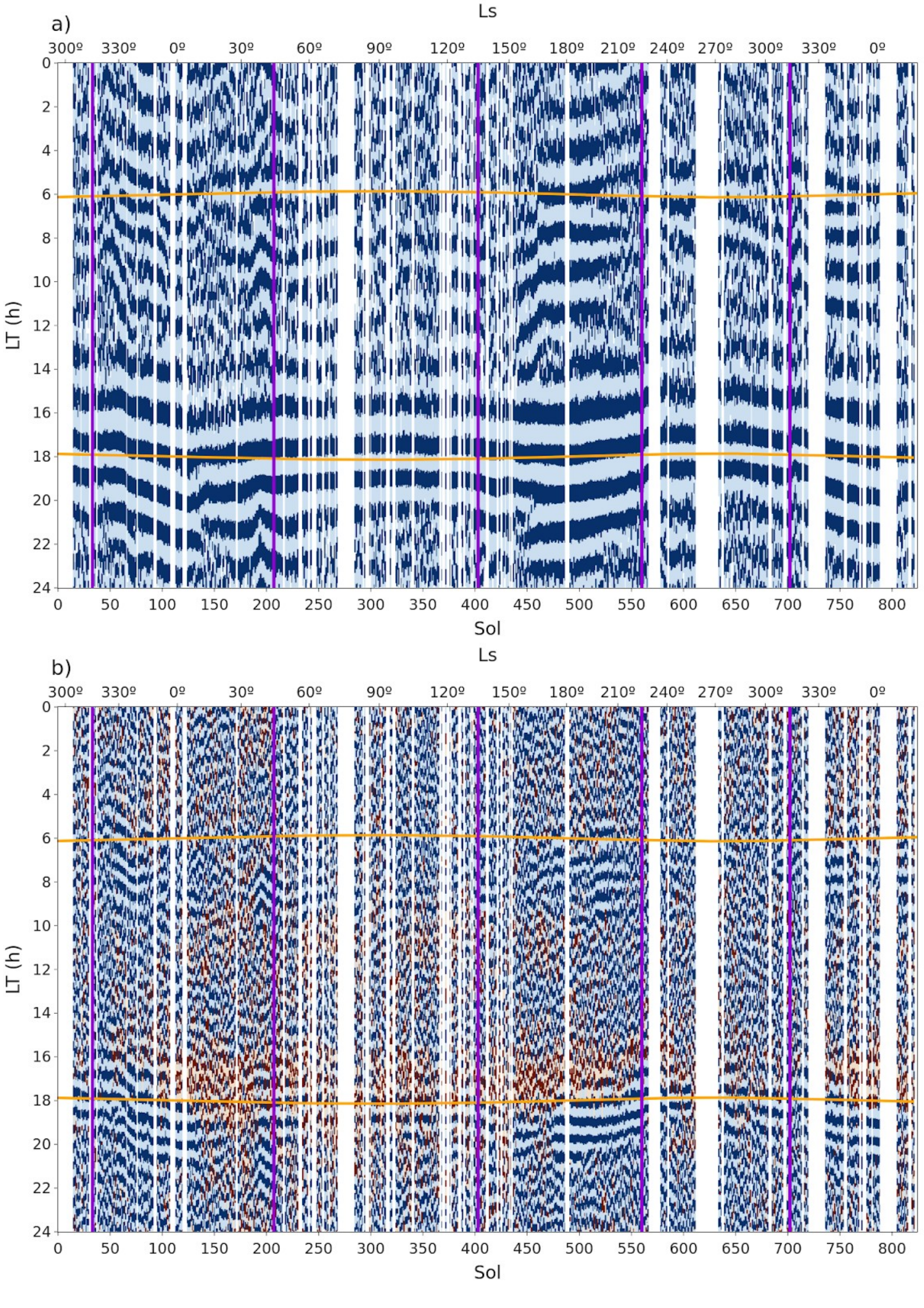

**Figure 5**. Visualization of the coherence of oscillations with periods (a) 3700-7400s and (b) 1800-3700s (given the applied bandpass filter, see text). Dark and pale blue colors indicate respectively a positive and negative anomaly of the bandpassed signal, equivalent reddish colors indicate oscillations below noise level. Horizontal stripes indicate the sol-to-sol coherency of oscillations. Other aspects of the graph (season, sunrise and sunset curves) are analogous to fig. 2a. Supporting figures S1 are equivalent to fig. 5 for spectral ranges 100-900s and 900-1800s.

In order to perform a systematic evaluation of the sol-to-sol coherence of pressure oscillations, we apply a bandpass filter to the signal corresponding to each sol, and plot the sign of the resulting pressure anomaly in a Ls-LT graph (the process is illustrated in supporting figure S2). Fig. 5 presents two cases of such a graph. Fig. 5a shows oscillations with periods in the range 3700-7400s, a range in which we can expect high-order harmonics of thermal tides, according to HB24, the coherency of oscillations (apparent in these graphs in the form of horizontal stripes) at most local times during the whole year probably indicates the predominance of tidal harmonics in these oscillations. Fig. 5b is the same kind of figure for oscillations with periods in the range 1800-3700s, it shows that coherency of oscillations is widespread in the dataset also at these higher frequencies, and very often in coincidence with GW activity as revealed by fig. 2a. GWs with higher frequencies in fig. 2 are often associated with less coherency in fig. 5b.

The signal interpreted as a soliton by Banfield, Spiga et al. (2020; their fig. 4b), happened around 19 to 20 LT, from sol 60 to sol 80, and it can be traced in fig. 5 (the interactive version of this figure, available in the supporting material, is specially useful for this purpose).

The equivalent graphs to fig. 5 for periods 900-1800s, and 100-900s are available in the supporting material. All these graphs display some degree of coherence in pressure anomalies in various seasons and local times.

### 3.4 Coincidence between GWs and absolute pressure variations

In HB24, a climatology of the pressure cycle driven by the ensemble of tidal harmonics was presented (HB24, their fig. 3), based on the derivative of pressure over time (i.e. pressure variations). Here we show in fig. 6 that the climatology of GWs from fig. 2a exhibits some coincidences with that climatology of the pressure cycle.

The background of fig. 6 (extracted from fig. 3 in HB24 and their dataset S4) represents the derivative of pressure. The red stripes after sunrise and sunset represent a quick ascension of the pressure that usually takes place at these times of the diurnal cycle, as can be seen for example in fig. 1a around 8 and 20 LT. We use the term “tidal bumps” to refer to these peaks in pressure after the sunrise and sunset, analyzed by Wilson et al. (2017) and Yang, Sun, et al. (2023).

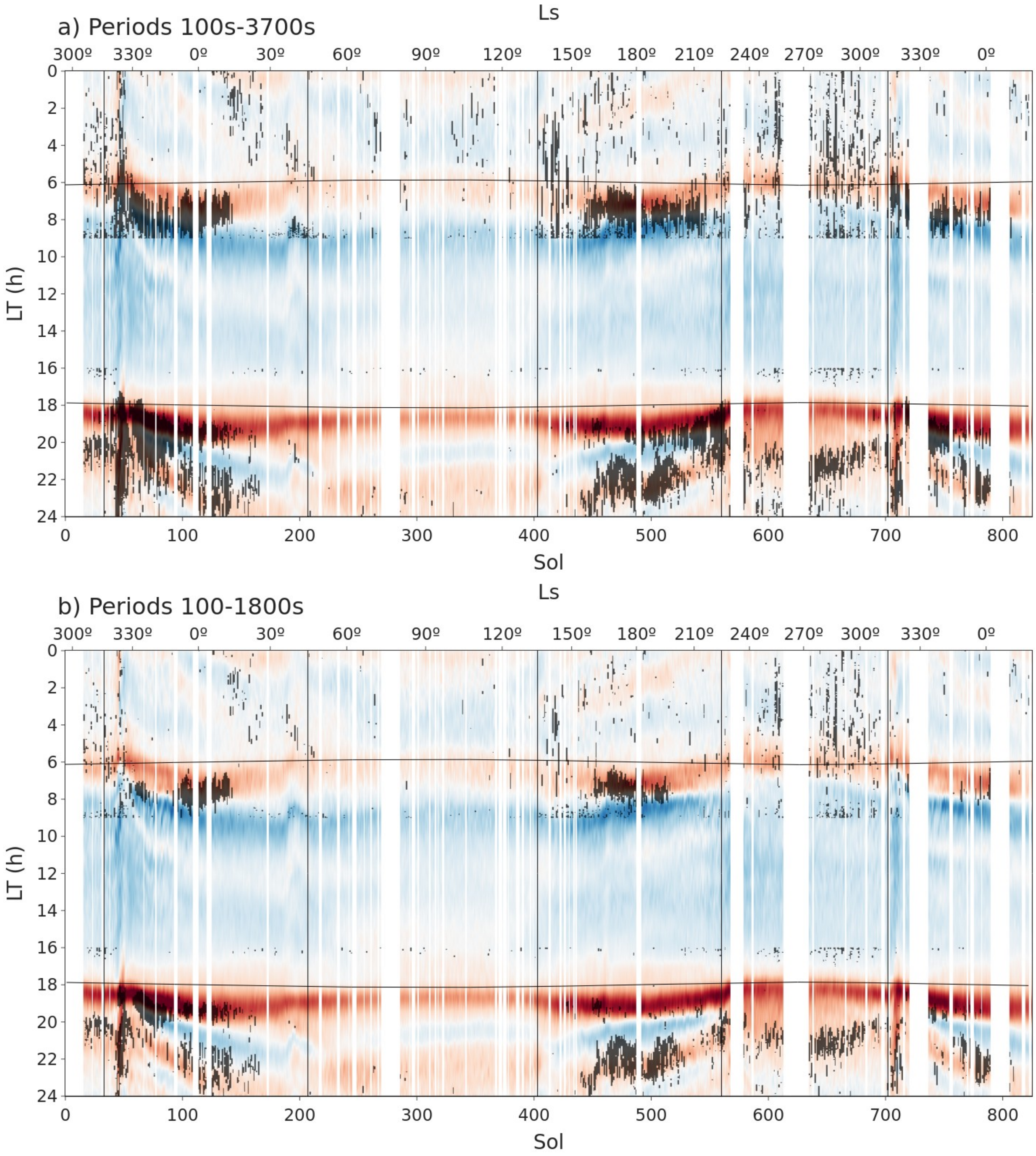

**Figure 6**. Coincidence between wavelet power from fig. 2a (highest wavelet power represented in black) and the diurnal pressure cycle, (represented by a background reproduced from fig. 3 and dataset S4 in HB24, where red represents pressure increment and blue represents pressure decreasement). a) Including wavelet power for periods 100s-3700s. b) Including wavelet power only for periods 100s-1800s. Wavelet power from 9 to 16 LT is not represented because it corresponds mostly to diurnal turbulence.

In front of this background, we superimpose the climatology of wavelet power previously presented in fig. 2a. The distribution of GWs in LT is well correlated to those moments of increasing pressure. The two stripes of evening GWs previously reported in section 3.1.2 are also well correlated to those seasons when two separate tidal bumps take place after sunset. Fig. 6b is equivalent to fig. 6a, but it only shows wavelet power corresponding to periods between 100s and 1800s, showing that the coincidence is still present for smaller periods.

This is a coincidence with a potential causal relation, but not a full correlation; wavelet power is not proportional to the pressure derivative, nor is it always larger when the derivative of pressure is positive or reaches some threshold. On the other hand, the pressure derivative is typically larger in equinox seasons, which are also the seasons when the coincidence is clearest.

In addition to this, some coincidence seems to be present also in the nocturnal wave activity (discussed in section 3.1.3) at the end of northward equinox season and starting of the southward equinox season, when the seasonal evolution of wavelet power mimics that of the red and blue stripes representing the pressure derivative.

## 4 Discussion

In section 3 we analyzed the climatology of GWs observed by InSight and showed some puzzling features: the sol-to-sol coherency (section 3.3), and the coincidence with pressure variations (section 3.4). In this section we provide some interpretations and discussions for these observations.

As explained in the introduction, the presence of GWs in a given location depends both on the local generation and propagation of GWs from remote sources. InSight is located in a flat area and we do not expect many orographic waves being generated locally, but other mechanisms (convection, wind shear) could produce waves locally. Observed waves can also be generated in remote locations, the nearby volcanoes of Elysium Montes are particularly likely candidates. The propagation of GWs can be a factor conditioning the observed climatology, and this propagation is largely affected by winds and by the thermal structure of the atmospheric column. We can imagine, for example, GWs propagating from longer distances at specific LTs and seasons as part of the cause for the observed climatology.

The Martian weather is known to be more repeatable compared to the Earth weather (for example, the daily repeating Arsia Mons Elongated Cloud; Hernández-Bernal et al., 2021a), and the sol-to-sol coherence of waves could be a manifestation of this repeatability. Daily repeating patterns in local winds (or more generally in large-scale and regional-scale atmospheric conditions) could induce GWs with similar phases in different sols. Such local winds could be ultimately caused by thermal tides, which might explain the coincidence described in section 3.4.

In addition to these interpretations based on the conventional knowledge about GWs on Mars, we propose two additional ideas that could also explain part of our observations.

### 4.1 The gray zone in the spectrum between GWs and thermal tides

Non-tidal GWs (transient local or regional oscillations, see introduction) with periods of several hours exist in the atmosphere of Mars. The finding of high-order tidal harmonics with periods of below two and even one Martian hour reported in HB24 implies that there is a “gray zone” in the spectrum of oscillations where both thermal tides and GWs could coexist. The range of the spectrum covered by this grayzone is unclear, it covers periods as long as a few hours, and could reach periods as short as less than one Martian hour if tidal harmonics beyond 24 are present in this atmosphere, as argued in HB24. The transient constructive interference of high-order harmonics to produce meteorologically relevant patterns that could be confused with GWs suggested in HB24 could happen within this gray zone of the spectrum.

The observation of oscillations with coherent phases in consecutive sols (section 3.3) and the coincidence between GWs and tidal bumps (section 3.4) could be to some extent symptoms of this gray zone. The coherence of waves observed here might actually be daily repeating patterns produced by thermal tides. And the coincidence with tidal bumps might be caused by a certain alignment of the phases of high-order and low-order tidal harmonics. However, high frequency GWs with periods much shorter than 1 hour apparent in figure 2b are less likely a result of high-order tidal modes, as they would require much higher harmonics.

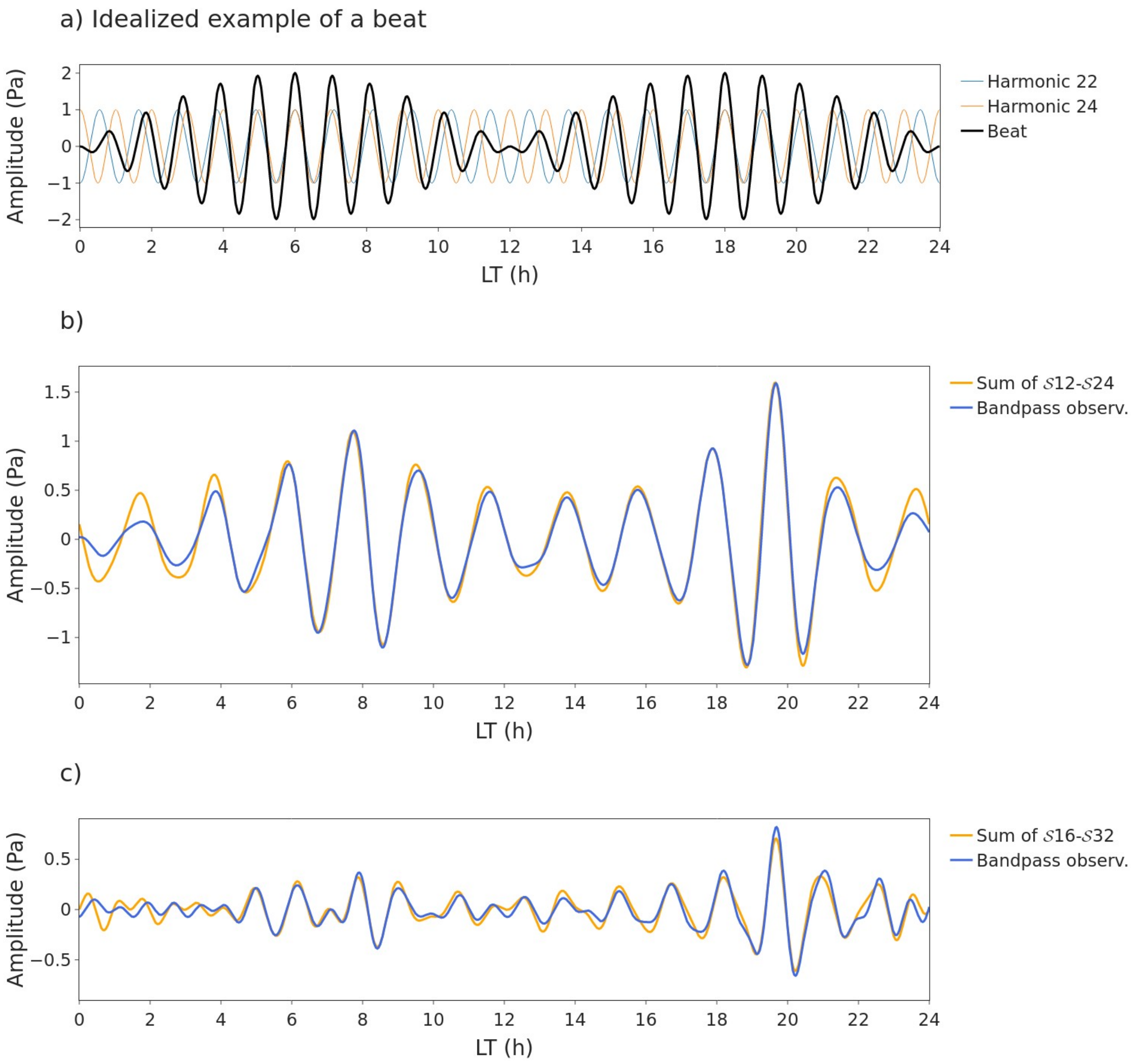

**Figure 7**. The interference of high-order harmonics. a) Example of a “beat”, which is a simple case of interference of two waves with similar periods. For this example, we choose the source waves to be harmonic numbers 22 and 24, with the same arbitrary amplitude. b) A real example of interference of many high-order harmonics (12-24) to reproduce wave patterns in sol 493 (previously shown in fig. 4). The observed signal has been bandpassed with a window of 3700-7400s (same spectral range as harmonics S12-S24). c) Same as b but for harmonics S16-S32 and corresponding bandpass window 2775-5550s (same spectral range as harmonics S16-S32). Note that the FFT is supposed to fit any signal, and therefore this figure is not a proof that these wave patterns are caused by tides. The only purpose of this figure is to illustrate that the interference of many harmonics can produce complex patterns.

The potential of high-order harmonics to interfere and produce complex patterns is illustrated in fig. 7. Fig. 7a shows a “beat”, which is a pattern produced by the interference of two waves with similar periods. This is a popular simple and idealized example of how different waves can interfere to produce complex patterns. Note that the apparent period of the resulting interference pattern is different from the period of the interfering harmonics, which points out how problematic the differentiation between non-tidal GWs and thermal tides in observed signals could be. Figs. 7b and 7c compare the interference of certain ranges of tidal harmonics from HB24 to the equivalent bandpass on sol 493 (also displayed in fig. 4). The intent of these panels is to show real examples of these patterns being caused by the interference of high-order harmonics of the solar rotation period, especially for the case of coherent waves (see the same sol 493 in fig. 4 and section 3.3). However, we must consider that, given its mathematical properties, the Fast Fourier Transform (FFT) used in HB24 to determine these tidal harmonics is expected to reproduce the input signal. Therefore the fact that we can reproduce the observed patterns as the interference of the harmonics is not unexpected and does not prove that these harmonics are the physical origin of those patterns.

### 4.2 Terminator Waves

Another explanation for the observation of GWs after the sunrise and sunset, and the coincidence with tidal bumps could be the generation of GWs by the movement of the terminator (Terminator Waves; TWs) previously mentioned in the introduction (where some relevant references were given). TWs are GWs generated around the terminator by the sudden change of insolation, and they are expected to take place after the sunset and the sunrise. A caveat of this explanation is the double band of GWs at different LTs observed in the evening in the equinox seasons (see section 3.1.2), which in principle cannot be explained in terms of TWs, and in general, the fact that the GW climatology in fig. 2a mimics the climatology of pressure derivative also in the night (section 3.4).

We note that the nature of TWs is sometimes unclear in the literature. Miyoshi et al. (2009) concluded that the TWs in their GCM “consisted of a superposition of the migrating tides s=4, 5, and 6”; to our understanding, this suggests that TWs might be sometimes being confused with thermal tides on Earth, which is actually analogue to the idea proposed in HB24 and explored here in section 4.1 that high-order tidal harmonics could produce transient patterns looking like GWs. Elucidating whether TWs compatible with our observations could exist in the lower atmosphere of Mars is a task for future works.

## 5. Conclusions

Based on the unprecedented dataset produced by the PS on Insight, we compiled a comprehensive climatology of GWs in the lower atmosphere of Mars and analyzed some features of those GWs. We used the non-conventional definition of seasons introduced in HB24 (section 2.2) and it turned out to be convenient to describe our results (fig. 2, and section 3).

The main result of this climatology is that most GW activity takes place after the sunrise and the sunset. The remarked dependence on LT (sunrise and sunset) is particularly interesting because previous works on GWs on Mars did not report diurnal variations (e.g. Kuroda et al. 2016; Heavens et al., 2020; Guzewich et al., 2021, is an exception).

GW activity undergoes a calm season at Ls 20-155º (section 3.1), which is the season of the MY with the lowest atmospheric dust content. Finding an explanation for this in future work would also be useful for interpreting other mesoscale/microscale phenomena occurring in the evening/morning and undergoing similar seasonal variability such as e.g. nocturnal turbulence (Chatain et al., 2021; Pla-Garcia et al., 2023).

Response to dust events (specially the annually repeating dust event C) and interannual variations have been observed (section 3.2). We showed the sol-to-sol coherency of waves (section 3.3) and the coincidence between the occurrence of GWs and the tidal increase of absolute pressure(section 3.4).

We discussed the overlap between GWs and thermal tides (section 4.1), an intriguing path opened by us in HB24 and explored here. We argue the existence of a “gray zone” in the spectrum of atmospheric oscillations on Mars, extending from periods of a few hours to potentially periods below one hour. The idea that some of the GWs observed here might actually be high-order thermal tides transiently interfering constructively provides a potential explanation to some features observed here, especially the sol-to-sol coherency of GWs, and maybe the coincidence of GWs and tidal bumps. But this only applies to those GWs with large periods, and cannot explain the existence of this coincidence also at higher frequencies (e.g. periods below 1800s in fig. 6b). Further observational and modelling efforts will be required to acquire a better understanding of the gray zone between non-tidal GWs and thermal tides.

Alternatively, these features and the observed climatology might be explained by more conventional mechanisms, connected to wind patterns and the propagability of GWs through the atmosphere, which would be ultimately caused by the diurnal cycle of thermal tides and the well known repeatability of Martian weather compared to Earth's one (section 4.2). In addition to it, in section 4.2 we suggested another non-tidal mechanism consistent on the so-called Terminator Waves (TWs), which have been investigated by various authors on Earth (Forbes et al., 2008; Miyoshi et al., 2009; Hedlin et al., 2018; see further details in the introduction), and have been modeled in the atmosphere of Mars (Forbes & Moudden, 2009). TWs could be an intriguing explanation to at least a part of the climatology that we observe in this work, as they are expected to happen after the sunrise and the sunset.

In practice, the interference of high-order tidal harmonics, the TWs, the natural repeatability of the Martian weather, and possibly other unidentified mechanisms, might play together to produce these observed features and the climatology reported in section 3.

Interestingly, other works on the Martian atmosphere show that rich dynamics take place on Mars around twilight (e.g. Connour et al., 2020; Hernández-Bernal et al., 2021a; Hernández-Bernal et al., 2021b). The findings about GWs around twilight reported in this work keep suggesting that many aspects of this moment of the Martian diurnal cycle remain unexplored.

A precedent to our climatology is the one compiled by Guzewich et al. (2021) based on MSL/REMS (Gómez-Elvira et al., 2012) pressure timeseries. Their results are overall reminiscent of ours, which suggests that these trends are not so much due to local mesoscale effects (as could be reasonably expected in the case of MSL/REMS, given that this mission is located inside the Gale crater), but they are connected to more global effects. InSight and MSL/REMS are both close to the equator of Mars and separated by only a few hundred kilometers, therefore further inSights into the detailed global climatology of GWs in the lower atmosphere will need to come from other landers (Perseverance, Zhurong, and any other future missions; Rodriguez-Manfredi et al., 2021; Li et al., 2021;) and from modeling.

As we learn more about the climatology of GWs in the lower atmosphere of Mars, the question of how it relates to the corresponding climatologies at other altitudes arises (a question addressed on Earth for example by Groot-Hedlin et al., 2017). Such analysis will require further efforts both from modeling and from other landers and satellites, among which non-sun-synchronous orbiters could play a central role, given the intriguing dependencies on LT reported here.

## Acknowledgments

This is InSight contribution number 348. The authors acknowledge the funding support provided by Agence Nationale de la Recherche (ANR-19-CE31-0008-08 MAGIS) and CNES. All co-authors acknowledge NASA, Centre National d'Études Spatiales (CNES) and its partner agencies and institutions (UKSA, SSO, DLR, JPL, IPGP-CNRS, ETHZ, IC, and MPS-MPG), and the flight operations team at JPL, CAB, SISMOC, MSDS, IRIS-DMC, and PDS for providing InSight data. The members of the InSight engineering and operations teams made the InSight mission possible and their hard work and dedication is acknowledged here.

We acknowledge Ralph Lorenz and an anonymous reviewer for their constructive comments.

## Open Research

The dataset of wavelet power and supporting videos and interactive figures are available in Hernández-Bernal & Spiga (2024).

InSight source data used in this study are publicly available in the Planetary Data System (Banfield, 2019).

For our analysis, we used the derived dataset Hernández-Bernal & Spiga (2023)

Python wavelet software provided by Evgeniya Predybaylo (2014) based on Torrence and Compo (1998).

Most recent dust climatologies (used for reference in fig. 2a) by Montabone et al. were obtained from:
https://www-mars.lmd.jussieu.fr/mars/dust_climatology/index.html

Figure 5 was made using the same procedure as figs. S1 in Hernández-Bernal et al. (2024). The background of figure 6 was made using a similar procedure to figure 3 of Hernández-Bernal et al. (2024). Figure 7 contains data about the tidal harmonics extracted from dataset S3 in the supporting materials of Hernández-Bernal et al., (2024).

# Supporting Information

**Contents of this file**

Text S1-S3

Figures S1

**Additional Supporting Information**

(Files available at https://doi.org/10.14768/ffb87d33-9694-4616-b89b-6833c18c8ee0 )

Caption for dataset S1

Caption for animations S1-S3, and S4

Captions for interactive figure S1

**Introduction**

This supporting material includes:

- Text S1. Details on the wavelet analysis
- Caption for dataset S1. 3D array containing compressed wavelet results.
- Text S2. List of sols removed for wavelet analysis
- Text S3. Details on normalization of wavelet power
- Caption for animations S1, S2,  S3. Showing the bandpassed timeseries for different sols and local times.
- Caption for interactive figure S1 (corresponding to fig. 2. in the main text)
- Caption for animation S4. Like fig. 2a for unfolded frequencies.
- Figures S1. Equivalent to fig. 5 for spectral ranges 100-900s and 900-1800s.

**Text S1. Details on the wavelet analysis.**

Our wavelet analysis is based on Torrence and Compo (1998). We use the python version of their implementation (Predybaylo, 2014) publicly available at:

https://github.com/ct6502/wavelets/tree/main/wave_python.

We tested the performance of different mother functions (wavelet functions), both on real data and on idealized synthetic signals. And we concluded that MORLET with param=6 (default value, which is automatically set if param=-1) works well in our case and produces a more accurate determination of the frequencies. We advise other researchers willing to use this technique to test different functions depending on the characteristics of their datasets. Wider wavelet functions result in wider COIs (Cone of Influence), which is manageable in our case because we count with very long continuous timeseries, but a more accurate selection of frequencies. Shorter wavelet functions (like PAUL) result in a less accurate selection of frequencies but a narrower COI, and it might be more suitable to analyze shorter timeseries produced for example by MSL/REMS.

We compute the wavelet power for the 24h of every sol, using 4h margin before and after the starting and the ending of the sol, this is to remove any effects from the COI, and it implies that some sols for which the whole 24h are available but the 4h margin is not available are removed from the dataset, this list of removed sols is available in text S2.

The computation is done over the derived dataset presented in HB24, which is downsampled to 1Hz. For storing and plotting we downsample the wavelet power to one sample per Martian minute (1440 samples per Martian sol). This wavelet power dataset is available in dataset S1.

**Caption for dataset S1. 3D array containing compressed wavelet power.**

This dataset is stored in npy format. In python in can be opened with:

```
import numpy as np
dataset=np.load('Dataset S1. Wavelet Power.npy')
```

The shape of the array stored in the file is: dataset.shape=(51,825,1440)

The first dimension (51) corresponds to the period (see below for the values of the periods), the second (825) dimension corresponds to the sol number (item number 100 corresponds to sol 100), and the third dimension (1440) corresponds to the local time, from 0h to 24h. Sols without data contain NaN everywhere.

The period (in seconds) corresponding to each one of the 51 layers in the first dimension is stored in a different npy file:

```
periods=np.load('Dataset S1. Periods.npy')
```

**Text S2. List of sols removed for wavelet analysis**

Our analysis is based on the derived dataset produced in HB24. The coverage of this dataset, its short interpolated gaps, and the number of full sols available were given in supporting tables S1, S2, S3 in HB24. Given the need for a margin of 4 hours imposed here for the computation of wavelet power, the actual list of sols available in dataset S1 is smaller. The original dataset contains 630 full sols, here we count with 615 sols. The list of removed sols is: 234, 240, 255, 298, 316, 318, 323, 340, 411, 436, 638, 684, 770, 775, 824.

**Text S3. Details on normalization of wavelet power**

As explained in the main text, the amplitude of wavelet power tends to be higher for higher periods. In many representations we are interested in putting together different periods, and this implies that a simple sum over all periods would result in a strong overrepresentation of high periods over low periods.

The average value of wavelet power for each period can be computed, from the array obtained in text S1, as:

```
A=np.nanmean(np.nanmean(dataset,axis=1),axis=1)
```

And this graph shows the result of this computation:

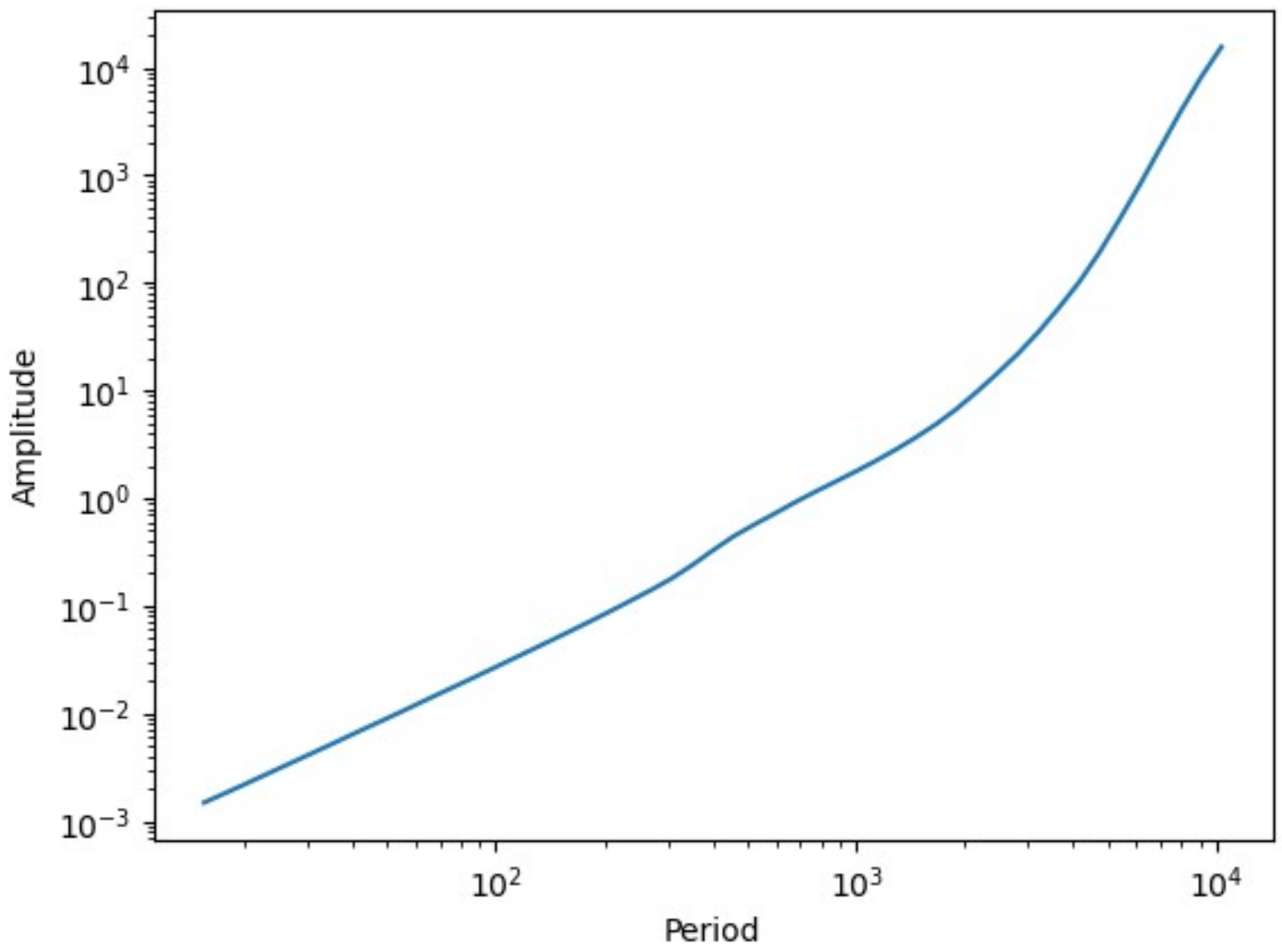

For the purpose of plotting together wavelet power corresponding to different periods, we divide values of wavelet power by the average for their period. This implies that we use arbitrary adimensional units in those plots. This normalization can be computed as:

```
normalized=np.zeros(dataset.shape)
for i in range(len(dataset)):
normalized[i]=dataset[i]/np.nanmean(dataset[i])
```

**Caption for animations S1,S2, S3. Bandpassed timeseries for different sols and LTs**

These animations are available in the files called:

- animS1.bandpass_100-3700s_00-06LT.gif
- animS2.bandpass_100-3700s_04-10LT.gif
- animS3.bandpass_100-3700s_18-24LT.gif

They are gif files containing for each sol a graph with the bandpassed timeseries (from 100s to 3600s). In each case, the horizontal axis is LT and the vertical axis is pressure anomaly (Pa). These timeseries can be compared to the computed wavelet power. A sample can be found in the image below.

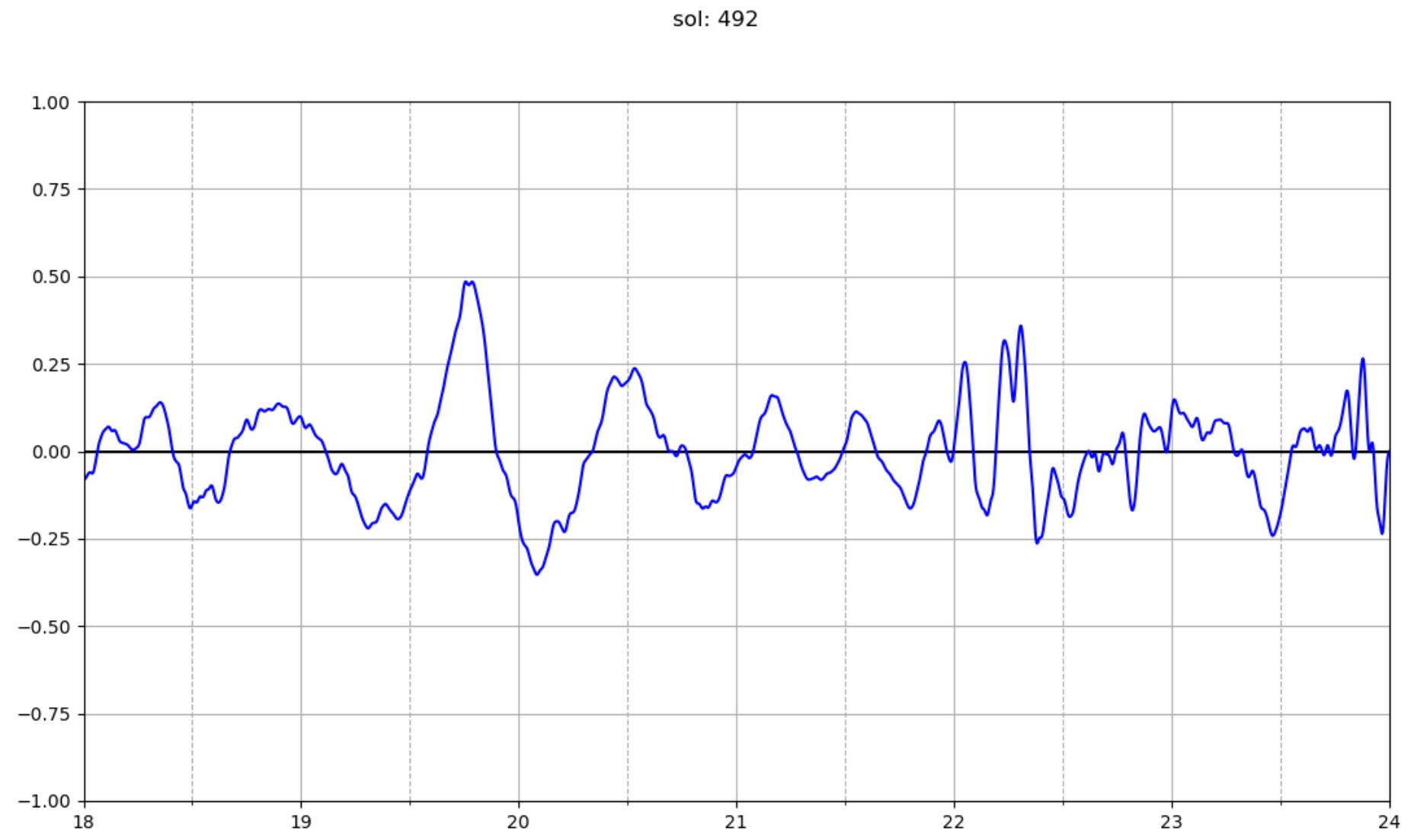

**Caption for interactive figure S1 (fig. 2 in main text)**

This interactive version of figure 2 can be found in two different files:

- Panel a: InteractiveS1_2a.html
- Panel b: InteractiveS1_2b.html

These interactive figures have been produced using the plotly library ( https://plotly.com/python/ ) and they can be visualized in any modern web browser.

**Caption for animation S4. Unfolding by frequencies of fig. 2a.**

Fig. 2a presents a normalized sum over all frequencies, as detailed in section 2.3.2 in the main text and in supporting text S3. This animation shows the same figure for only specific frequencies. It can be found in the file named: animS4.fig2a_unfolded.gif And it can be visualized by frames for example using the gifview app in linux. Find below two samples from the animation.

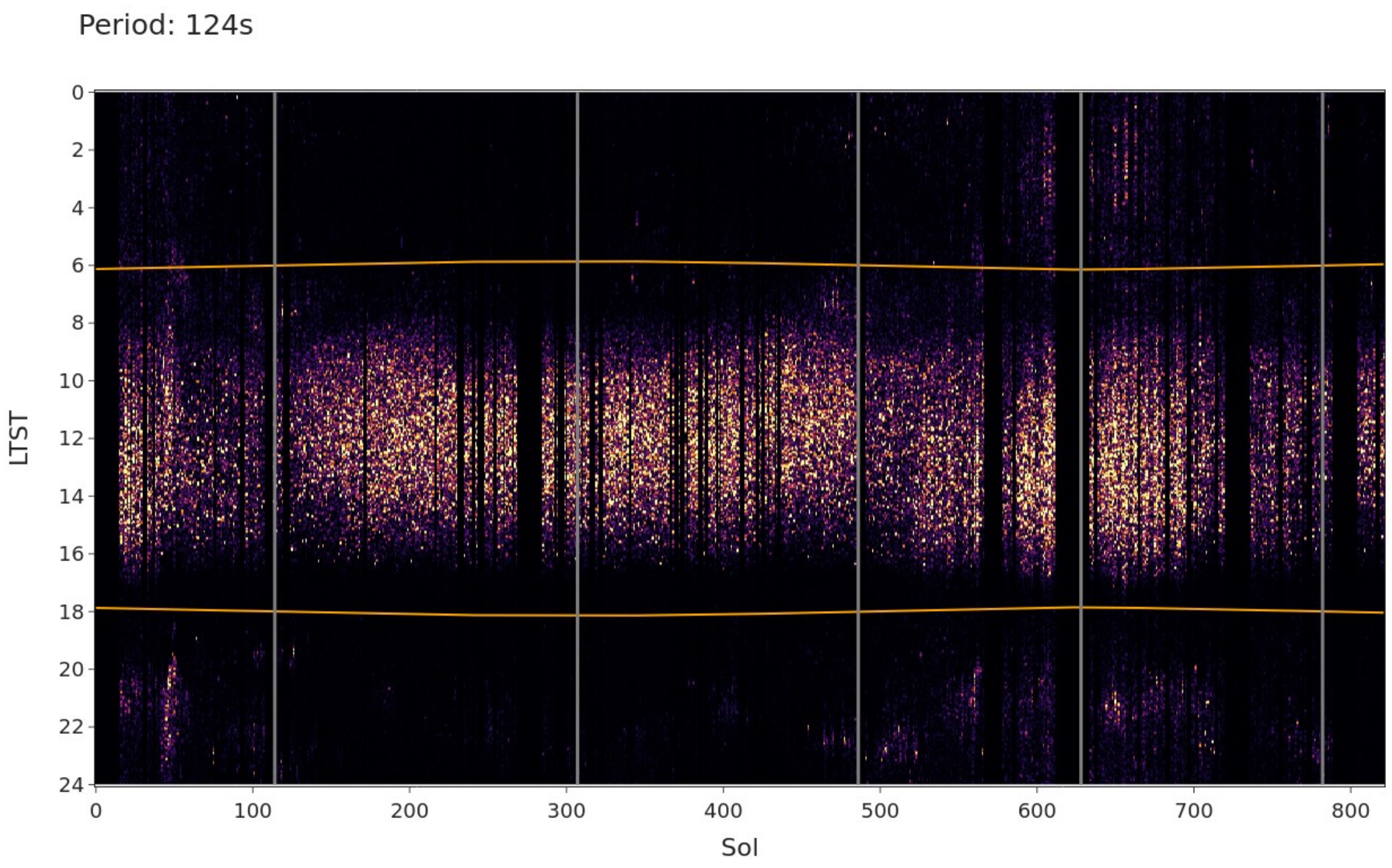

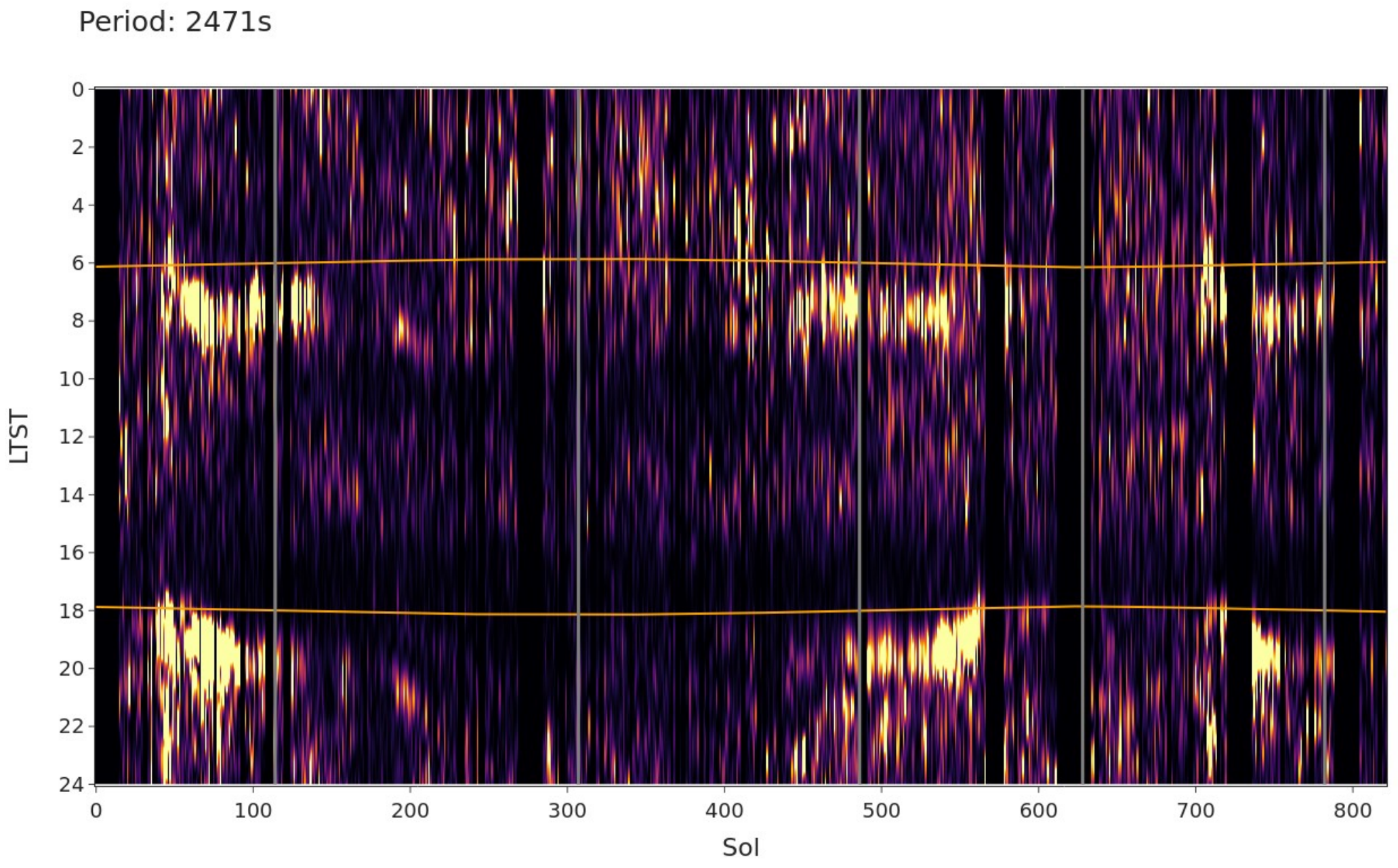

**Figures S1. Equivalent to fig. 5 for spectral ranges 900-1800s and 100-900s.**

These figures are equivalent to fig. 5 but for different spectral ranges. They are intended to show that even at periods below half an hour there is some level of coherency in the pressure anomalies from one sol to the next.

**Figure S1a**

Bandpass at 900-1800s

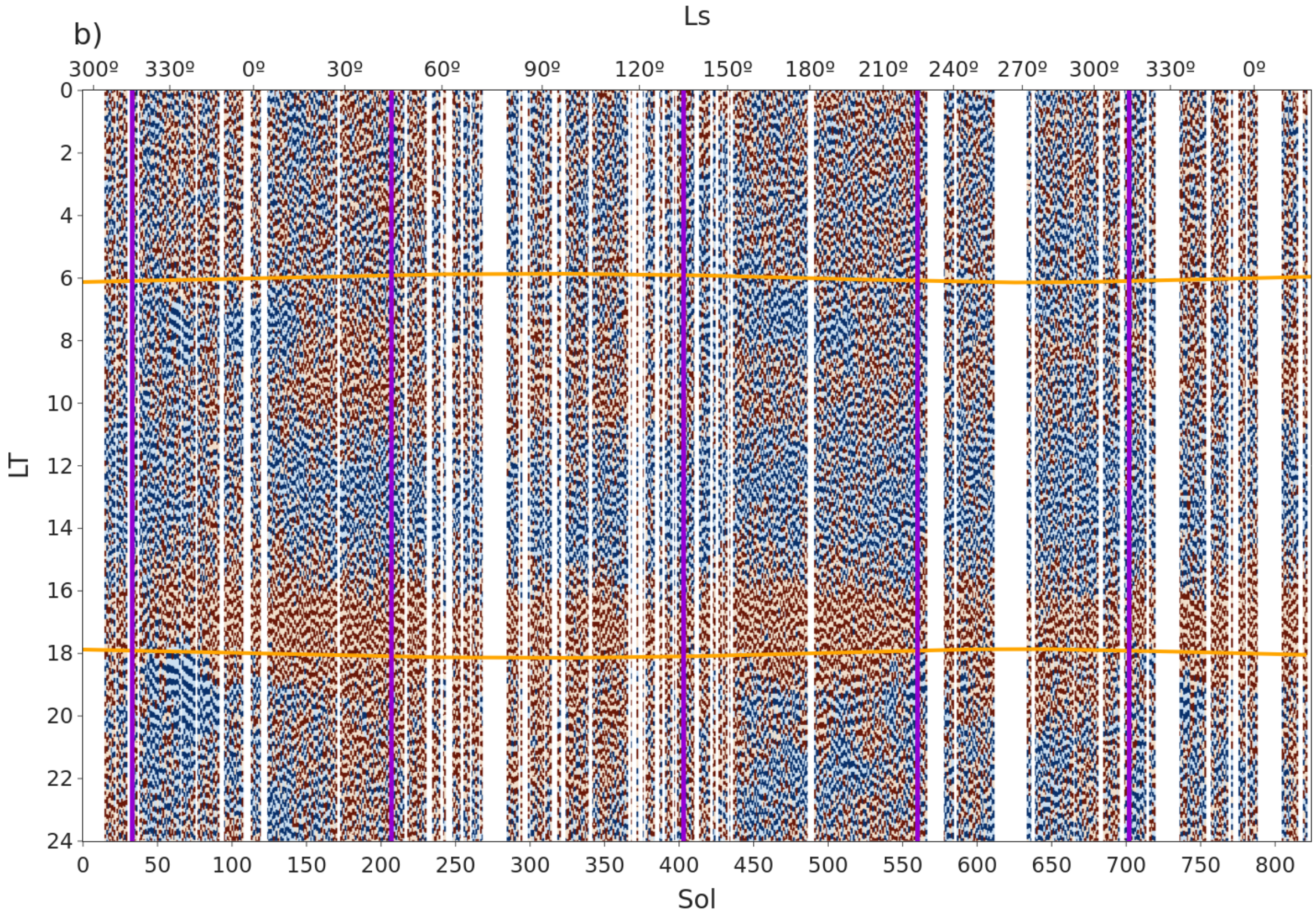

**Figure S1b**

Bandpass at 100-900s

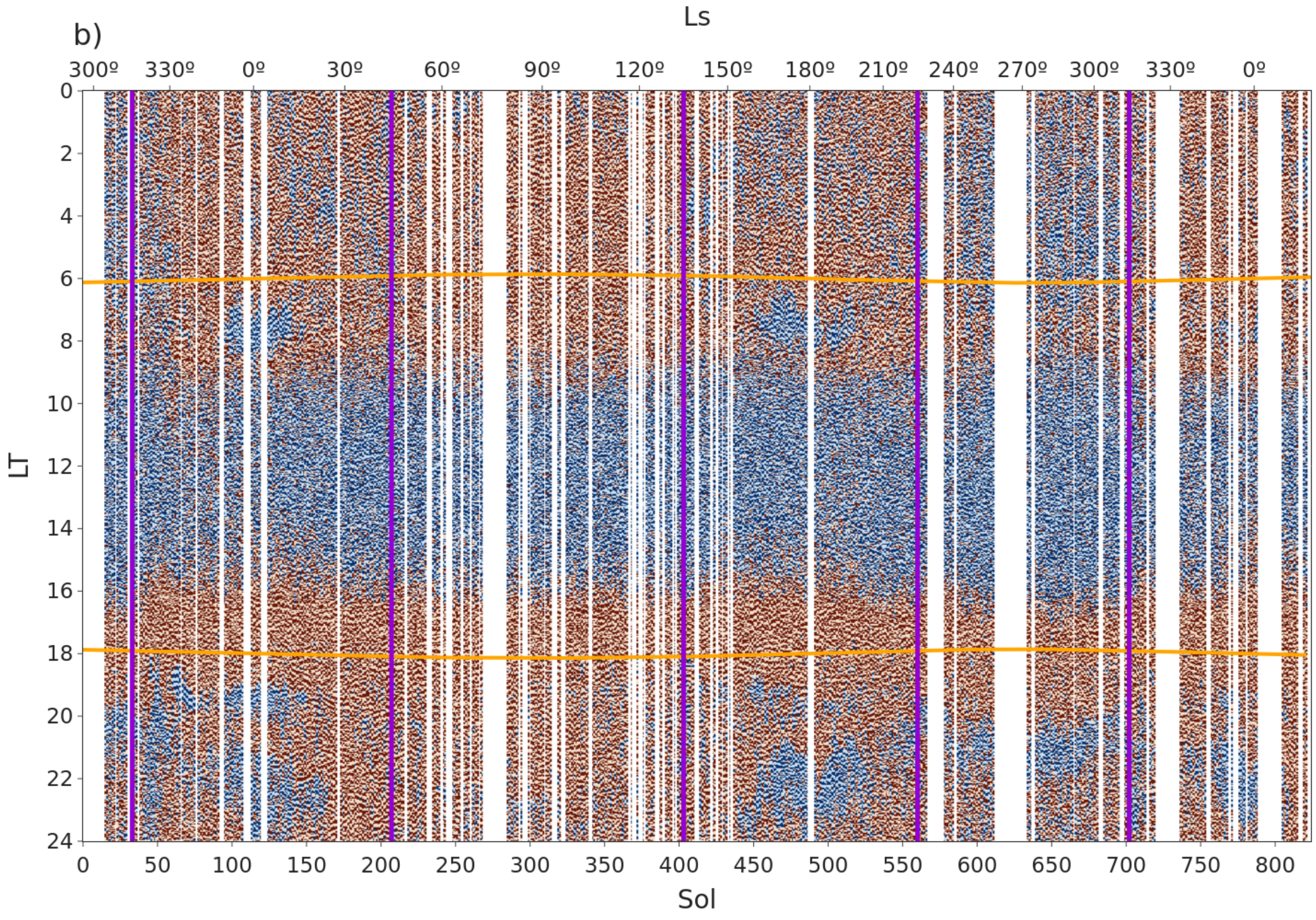

**Figures S2. Example of how fig. 5 and figs S1 are made**

Figs. 5 and S1 represent the sign of the pressure anomaly as a function of sol and LT. For clarity, we show here an example of how these figures are made. We show the pressure anomaly for sol 156 after a bandpass for periods from 1800 to 3700s. Positive anomalies are shadowed in dark blue, negative anomalies are represented in pale blue. Those colors are then translated for every sol into figs. 5 and S1. This is the same procedure employed in HB24, also explained in the supporting material there.

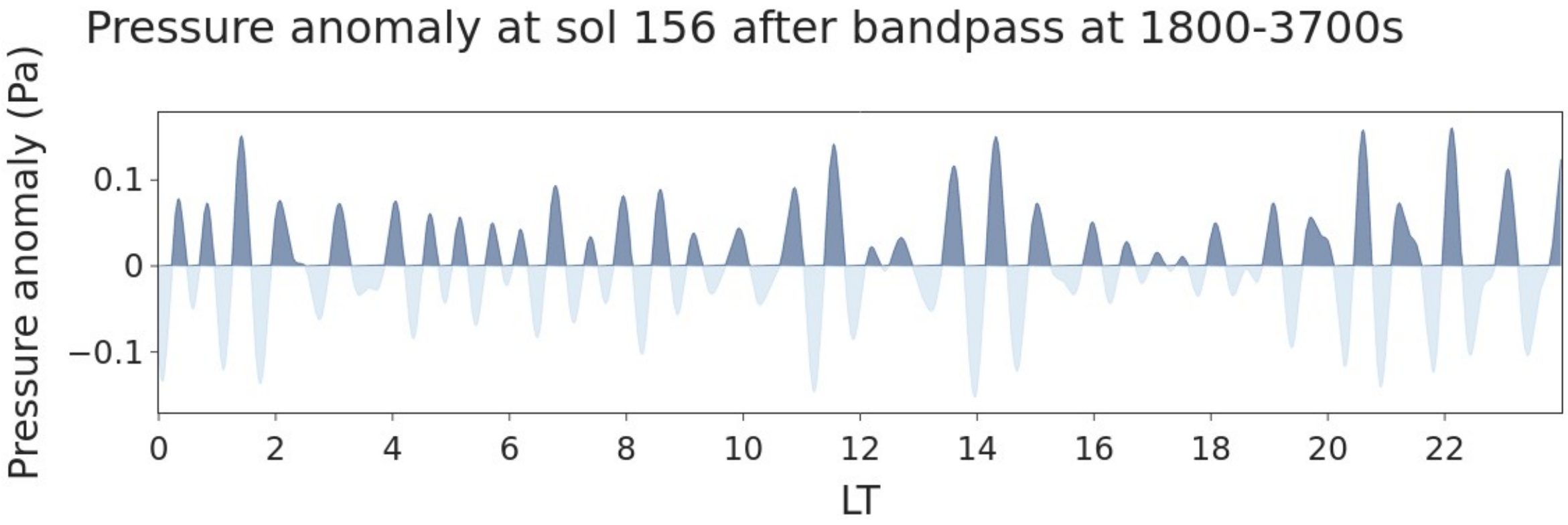